\documentclass[11pt]{article}
\usepackage{amsmath,amssymb,amsfonts,amstext}
\usepackage{tabularx}
\usepackage{hyperref}
\usepackage{epsf}
\usepackage{graphicx}
\usepackage{amsfonts}
\usepackage{subfigure}
\usepackage[usenames]{color}

\definecolor{Blue}{rgb}{0,0.08,0.45}
\definecolor{Magenta}{cmyk}{0.1,0.8,0,0.1}
\definecolor{Orange}{rgb}{1,0.5,0}
\begin{document}

\title{\hfill\textbf{\small CERN-PH-TH/200-022}\\ \hfill\textbf{\small OUTP-10-05P}\\ 
Hybrid Natural Low Scale Inflation}
\author{ Graham G Ross \\
{\normalsize \textit{Rudolf Peierls Centre for Theoretical Physics,} }\\
{\normalsize \textit{University of Oxford, 1 Keble Road, Oxford, OX1 3NP, UK}}
\\
{\normalsize \textit{and} }
\\
{\normalsize \textit{Department of Physics, CERN - Theory Division,} }\\
{\normalsize \textit{CH-1211 Geneva 23, Switzerland} }\\
\\
Gabriel Germ\'an\\
{\normalsize \textit{Instituto de Ciencias F\'isicas} }\\
{\normalsize \textit{Universidad Nacional Aut\'onoma de M\'exico,}}\\
{\normalsize \textit{Apdo. Postal 48-3, 62251 Cuernavaca, Morelos, Mexico}}}
\date{}
\maketitle

\begin{abstract}
We discuss the phenomenological implications of hybrid natural inflation models in which the inflaton is a pseudo-Goldstone boson but inflation is terminated by a second scalar field. A feature of the scheme is that the scale of breaking of the Goldstone symmetry can be lower than the Planck scale and so gravitational corrections are under control. We show that, for supersymmetric models, the scale of inflation can be chosen anywhere between the Lyth upper bound and a value close to the electroweak breaking scale. Unlike previous models of low scale inflation the observed density perturbations and spectral index are readily obtained by the choice of the free parameters.
\end{abstract}

\section{Introduction}

Even though the concept of inflation was proposed
almost thirty years ago \cite{Guth:1981}, \cite{Linde:1982}, \cite%
{Albrecht:1982} only a handful of models implementing it can be considered
to have a compelling physical basis. Essentially the problem arises due to
the fact that the inflaton has to be a very light field with mass less than
the Hubble expansion parameter during inflation. Quantum or gravitational
contributions to the inflaton mass are typically too large unless there is
an underlying symmetry protecting the inflaton from such corrections.
Although supergravity can protect against large radiative contributions the gravitational corrections are typically of order the Hubble
scale and too large (the \textquotedblleft eta\textquotedblright  problem); only a Goldstone (shift) symmetry can avoid such
contributions. In this case the inflaton is a pseudo Goldstone boson, its
small mass being due to the breaking of the continuous Goldstone symmetry by
an anomaly at the quantum level or by explicit breaking.

The original inflationary model based on a pseudo Goldstone inflaton was the  \textit{Natural Inflation} model of
Freese \textit{et al.} \cite{Freese:1990rb,Freese:1993,Freese:2008if}. It was based
on an anomalous Abelian symmetry such that  the quantum anomaly generated mass for the would-be Goldstone mode. A potential problem for the model is due to the
fact that, in order to generate sufficient inflation, the scale of symmetry
breaking must be bigger than the Planck scale and in this case there may be
large Quantum Gravity corrections of $O(f/M_{Planck}).$ \footnote{In models with
additional scalar fields it is possible to avoid such super Plankian scales%
\cite{Kim:2004rp,Dimopoulos:2005ac} while maintaining the \textit{Natural Inflation} form of the inflaton potential with an effective super Plankian scale $f.$}

Models of \textit{Natural Inflation} with explicit breaking of the Goldstone symmetry
have received relatively little attention. Small terms explicitly breaking
the continuous symmetry arise if the underlying symmetry is a discrete
symmetry that gets promoted to a continous one if only the low dimension
(renormalisable) terms are kept in the Lagrangian. In this case the higher
dimension terms explicitly breaking the continuous symmetry are suppressed
by a large inverse mass scale and are naturally small. In these models to avoid the
need for a super-Plankian scale, $f,$ the end of inflation must be triggered
by a second scalar field \cite{Linde:1994}. In this case
the number of e-folds of inflation depends on an additional parameter
allowing for viable models with a sub-Plankian value for $f$ \cite{Shaun:2008}. It proves to
be impossible to arrange for such hybrid natural inflation based on an Abelian
global symmetry but models of hybrid natural inflation have been developed
with an underlying non-Abelian discrete symmetry \cite{Stewart:2000pa,Graham:2009}. In this letter we explore the phenomenology of these models
in detail, concentrating on the possibility that the scale of inflation may be very low. To achieve low scale inflation the hybrid field (which is not a pseudo Goldstone boson) must be light and this in turn requires that the underlying theory should be supersymmetric to protect the mass of the hybrid field from large radiative corrections. 

We start with a model independent discussion of the structure and phenomenological implications of hybrid natural inflation before turning to a more detailed discussion of a particular supersymmetric model.

\section{Slow-roll parameters and observables}

As discussed in \cite{Graham:2009}, the terms of the (hybrid) inflaton
potential relevant when density perturbations are being produced have a
simple universal form corresponding to the slow roll of a single inflaton
field $\phi $:%
\begin{equation}
V\simeq V_0\left( 1+a\cos \left( \frac{\phi }{f}\right) \right) ,
\label{potential}
\end{equation}%
where $a$ is a constant. \textit{Natural Inflation} corresponds to the case $a=1$
while \textit{Hybrid Natural Inflation} has $a<1.$ We first list the detailed
expressions for the observables, the spectral index $n_{s},$ the density
perturbation at wave number $k$, the tensor to scalar perturbations ratio, $%
r,$ and the \textquotedblleft running\textquotedblright\ index $n_{r}.$
These are described in terms of the scale $\phi _{\mathrm{H}}$ at which the
perturbations are produced, some $40- 60$ e-folds before the end of
inflation, together with the scale of inflation given in terms of $V_{0}$
and the parameters $a$ and $f.$ We will discuss later the model constraints
on these parameters.  \textit{Hybrid Natural Inflation} is consistent with recent observational bounds coming from the five years data of the WMAP team \cite{Komatsu:2008}, \cite{Dunkley:2008} and those from BAO \cite{Percival:2007} and SN surveys \cite{Riess:2004}, \cite{Astier:2006}, \cite{Woods-Vasey:2006}. A Table displaying these results can be found in \cite{Graham:2009}.

The observables are given in terms of the usual slow-roll parameters \cite%
{Liddle:2000cg}.

\begin{eqnarray}
\epsilon &\equiv & \frac{M^{2}}{2}\left( \frac{V^{\prime }}{V}\right) ^{2}\simeq%
\frac{1}{2}M^{2}\frac{a^{2} }{f^{2}}\sin ^{2}\left( \frac{\phi }{f}\right)%
,\qquad \\
\eta &\equiv &M^{2}\frac{V^{\prime \prime }}{V}\simeq-M^{2}\frac{a }{f^{2}}\cos \left( 
\frac{\phi}{f}\right), \\
\xi &\equiv &M^{4}\frac{V^{\prime }V^{\prime \prime \prime }}{V^{2}}\simeq-M^{4}%
\frac{a^{2} }{f^{4}}\sin ^{2}\left( \frac{\phi}{f}\right)=-2\left( \frac{M}{f}\right)^2\epsilon,
\end{eqnarray}%
where $M$ is the reduced Planck scale, $M=2.44\times 10^{18}$~GeV.

In terms of these the observables are given by 
\begin{eqnarray}
r &=&16\epsilon , \\
n_{\mathrm{s}} &=&1+2\eta -6\epsilon ,  \label{spectral} \\
n_{\mathrm{r}} &=&16\epsilon \eta -24\epsilon ^{2}-2\xi, \label{run} \\
\delta _{\mathrm{H}}^{2}(k) &=&\frac{1}{150\pi ^{2}}\frac{V_{\mathrm{H}}}{%
\epsilon _{\mathrm{H}}M^{4}} .
\end{eqnarray}

\subsection{\textit{Natural Inflation.}} 

For the case of \textit{Natural Inflation} $\phi $ is the only scalar field and the
end of inflation is determined by the potential of Eq.(\ref{potential}) with $%
a=1.$ Thus the inflaton value, $\phi _{e},$ at the end of inflation is
determined and, up to an uncertainty about the number of intermediate
e-folds of inflation, hence so too is the inflaton value, $\phi _{\mathrm{H}}$, at
the time the density perturbations relevant today leave the horizon. As a
result \textit{Natural Inflation} has only two free parameters, the scale of inflation $\Delta $ \ and $f$.
The spectral index is always less than one and using the WMAP 3-year data
gives the bound \cite{Freese:2008if}  $f>3.5M$. For $f$ close to this bound
the density fluctuations constrain the scale of inflation to lie in the
range $\Delta =10^{15}GeV-10^{16}GeV.$ With this $r$ and $n_{r}$ are
determined 
\begin{equation}
r=4\left( \delta _{\mathrm{ns}}-\frac{M^{2}}{f^{2}}\right) ,
\label{indicesnatr}
\end{equation}%
\begin{equation}
n_{r}=-\frac{1}{2}\left( \delta _{\mathrm{ns}}^{2}-\frac{M^{4}}{f^{4}}%
\right) ,  \label{indicesnatnr}
\end{equation}%
where $\delta _{\mathrm{ns}}\equiv 1-n_{s}.$ This implies $r\leq 0.2$ and $%
\left\vert n_{r}\right\vert <10^{-3}$, where we have used the value $n_{s}\geq 0.947$ \cite {Komatsu:2008}, \cite {Dunkley:2008} .

\subsection{\textit{Hybrid Natural Inflation.}} 

For the case of \textit{Hybrid Natural Inflation} $\phi _{e}$ is determined by the
hybrid sector of the theory so $\phi _{e},$ or more conveniently $\phi
_{\mathrm{H}},$ is essentially a free parameter. Thus \textit{Hybrid Natural Inflation} has
four free parameters namely $\Delta $, $f$, $a$ and $\phi _{\mathrm{H}}$.  Using the
equation for the spectral index Eq.(\ref{spectral}) at $\phi _{\mathrm{H}}$
we find 
\begin{equation}
\delta _{\mathrm{ns}}=\frac{2a}{f^{2}}\frac{c_{\mathrm{H}}}{(1+ac_{\mathrm{H}%
})}+\frac{3a^{2}}{f^{2}}\frac{(1-c_{\mathrm{H}}^{2})}{(1+ac_{\mathrm{H}})^{2}%
},  \label{spectral2}
\end{equation}%
where $c_{\mathrm{H}}\equiv \cos \left( 
\frac{\phi _{\mathrm{H}}}{f}\right) .$ Then 
\begin{equation}
f^{2}\delta _{\mathrm{ns}}\equiv z=\frac{a(2c_{\mathrm{H}}+a(3-c_{\mathrm{H}%
}^{2}))}{(1+ac_{\mathrm{H}})^{2}} ,  \label{z}
\end{equation}%
showing that $f\varpropto \sqrt{a}$ for small $a.$

At $\phi _{\mathrm{H}}$ the solutions to Eq.(\ref{z}) are given by 
\begin{equation}
c_{1\mathrm{H}}=\frac{1-z+\sqrt{1+3a^{2}-3\left( 1-a^{2}\right) z}}{a\left(
1+z\right) },\text{\ }a\geqslant \frac{1}{3}  \label{solution1}
\end{equation}%
and 
\begin{equation}
c_{2\mathrm{H}}=\frac{1-z-\sqrt{1+3a^{2}-3\left( 1-a^{2}\right) z}}{a\left(
1+z\right) },\;a<1.  \label{solution2}
\end{equation}

The first solution in the limit $a=1$ corresponds to \textit{Natural Inflation}. However to avoid the possibility of large gravitational corrections to the
potential we will concentrate on the case $f<M.$ Then it follows from Eq.$%
\left( \ref{spectral2}\right) $ that, in order to obtain a small $\delta _{%
\mathrm{ns}},$ the parameter $a$ must be small ($a<0.026$ for $n_{s}\geq 0.947$) and so the
relevant solution is the second one.

We are now able to discuss the phenomenological implications of
\textit{Hybrid Natural Inflation}. Eq.$\left( \ref{z}\right) $ implies 
\begin{equation}
\delta _{\mathrm{ns}}\simeq 2a\left( \frac{M}{f}\right)^{2} c_{\mathrm{H}},
\label{fluctuations}
\end{equation}%
where now and in what follows $c_{\mathrm{H}}$ corresponds to $c_{2\mathrm{H}}$ of Eq.(\ref{solution2}).
The hybrid sector triggers the end of inflation through the growth of a term
proportional to $\sin \left( \phi /f\right) $ driving the mass squared of
the hybrid field to be negative. To avoid introducing a fine tuning between
terms in this sector it is necessary that $\sin \left( \phi /f\right) $
should be varying rapidly for $\phi \approx \phi _{e}$ corresponding to $%
\phi /f\ll \pi /2,$ $c_{\mathrm{H}}\simeq 1.$ Thus, $c.f.$ Eq.(\ref{fluctuations}), fitting the spectral index
essentially fixes the ratio $a/f^{2}$. The remaining observables are given by 
\begin{eqnarray}
r &=&\frac{8a^{2}(1-c_{\mathrm{H}}^{2})}{f^{2}(1+ac_{\mathrm{H}})^{2}} \simeq
4\,a\,\delta _{\mathrm{ns}}\,(1-c_{\mathrm{H}}^{2})\, \simeq 4\,a\,\delta _{\mathrm{ns}}\,\left( 
\frac{\phi _{\mathrm{H}}}{f}\right)^2  < 2\,\delta _{\mathrm{ns}}^{2}\,\left( 
\frac{f}{M}\right)^2 ,  \label{tensor} \\
n_{r} &=&\frac{2a^{2}(1-3a^{2}-2ac_{\mathrm{H}})(1-c_{\mathrm{H}}^{2})}{%
f^{4}(1+ac_{\mathrm{H}})^{4}}\simeq \frac{1}{2}\delta _{\mathrm{ns}%
}^{2}\,(1-c_{\mathrm{H}}^{2}) \, \simeq \frac{1}{2}{\delta _{\mathrm{ns}}^2 \left( 
\frac{\phi _{\mathrm{H}}}{f}\right)^2  < \frac{1}{2}\delta _{\mathrm{ns}}^{2}\, ,}  \label{running} \\
A_{\mathrm{H}}^{2} &= &75\pi ^{2}\delta _{\mathrm{H}}^{2}=8\frac{V_{\mathrm{H}}}{%
M^{4}}\frac{1}{r}\equiv8\left(\frac{\Delta}{M}\right) ^{4}\frac{1}{r}\, ,  \label{delta4}
\end{eqnarray}
where the observed magnitude of the density perturbations corresponds to $%
A_{\mathrm{H}}\,\equiv\, \sqrt{75}\,\pi\, \delta _{\mathrm{H}}$ and $\delta _{\mathrm{H}}\simeq \left(1.91\times 10^{-5}\right).$ From this we
see that the tensor to scalar ratio is small bounded by $r<5.6\times10^{-3}$ but is
typically much smaller since $c_{\mathrm{H}}\simeq 1$ and $a$ may be much smaller
than $0.026.$ The running index is bounded as $ n_{r}<1.4\times 10^{-3}$ and the scale of inflation is also bounded, $\Delta^{4}=\frac{1}{8}M^4A_{\mathrm{H}}^{2}r$ so $\Delta <9\times10^{15}\left( 
\frac{f}{M}\right)^{1/2} GeV.$ However much lower scales of inflation through the choice of $a\left( 1-c_{\mathrm{H}}^{2}\right) $ can be obtained. For
small $\phi _{\mathrm{H}}/f$ we have 
\begin{equation}
\Delta =M\sqrt{\frac{A_{\mathrm{H}}\delta _{\mathrm{ns}}}{2}\left( \frac{%
\phi _{\mathrm{H}}}{M}\right) }\simeq 9\times10^{15}\sqrt{ \frac{\phi _{\mathrm{H}}}{M} } \, GeV .  \label{scale}
\end{equation}
From this it is clear that to achieve low scale inflation it is necessary that $\phi_{\mathrm{H}}$ should be small. As mentioned above $\phi_{\mathrm{H}}$ is determined by the field value $\phi_e$ at the end of inflation and this is in turn determined by the hybrid sector of the theory. In non-supersymmetric hybrid models $\Delta \ge m_\chi$ in order to have zero cosmological constant after inflation and so low scale inflation requires low hybrid mass ($m_\chi$ is the mass of the hybrid field $\chi$ defined below). This in turn points to a supersymmetric model of inflation since the hybrid field, which is not a pseudo-Goldstone boson, needs supersymmetric protection against large radiative corrections to its mass (the hierarchy problem again). Thus we turn to a specific supersymmetric example of a hybrid sector in order to illustrate the expectation for the range of $\phi_e$ and hence of the scale of inflation.

\section{A supersymmetric example}

The field content of the supersymmetric model \cite{Graham:2009} consists of chiral supermultiplets of the $N=1$ supersymmetry which
transform non-trivially under a $D_{4}$ non Abelian discrete symmetry. The
supermultiplets consist of a $D_{4\text{ }}$doublet $\varphi =\left( 
\begin{array}{c}
\varphi _{1} \\ 
\varphi _{2}%
\end{array}%
\right) $ and three singlet representations $\chi _{1,2}$ and $A$
transforming as $1^{-+}$, $1^{--}$ and $1^{++}$ respectively where the superscripts refer to the transformation properties under the $Z_2$ semidirect product factors of $D_4$ ($D_4=Z_2\ltimes Z_2'$ ). The
interactions are determined by the superpotential given by%
\begin{equation}
P=A\left( \Delta ^{2}-\lambda _{3}\frac{1}{M^{2}}\varphi _{1}\varphi
_{2}\chi _{1}\chi _{2}\right) ,
\end{equation}%
where $\Delta$ is a constant with units of mass.
These are the leading order terms consistent with an additional $R-$symmetry
under which the fields $A,\varphi ,\chi $ have charges $2,-2$ and $2$
respectively.

The field $\varphi $ has the form 
\begin{equation}
\varphi =e^{i\mathbf{\phi \cdot \sigma }}\left( 
\begin{array}{c}
0 \\ 
\rho +v%
\end{array}%
\right) =\frac{\rho +v}{\phi }\left( 
\begin{array}{c}
\left( \phi _{2}+i\phi _{1}\right) \sin \left( \frac{\phi }{v}\right) \\ 
\phi \cos \left( \frac{\phi }{v}\right) -i\phi _{3}\sin \left( \frac{\phi }{v%
}\right)%
\end{array}%
\right) ,
\label{V}
\end{equation}%
where $\sigma _{i}$ are the Pauli spin matrices, $\phi _{i}$ are the pseudo
Goldstone fields. The field $\rho $ acquires a mass of $%
O(m_{\varphi })$ and plays no role in the inflationary era.

Inflation occurs for small $\frac{\phi _{i}}{v}.$
In this region, for positive $\lambda _{2},$ the field $\phi _{3}$ has a
positive mass squared while $\phi _{1,2}$ have negative mass squared. Thus $%
\phi _{3}$ does not develop a $vev.$ The full potential for the fields
acquiring $vev$s then has the form%
\begin{eqnarray}
V\left( \phi ,\chi \right) &\approx&\Delta ^{^{\prime }4}+64\lambda _{2}\frac{m^{2}%
}{M^{2}_M}f^{4}\cos \left( \frac{\phi }{f}\right)
+\sum\limits_{i=1}^{2}m_{\chi _{i}}^{2}\left\vert \chi _{i}\right\vert ^{2}-
\notag \\
&&-8\lambda _{3}e^{i\alpha }\Delta ^{2}\frac{f^{2}}{M^{2}}\sin \left( \frac{%
\phi }{2f}\right) \chi _{1}\chi _{2}+h.c.+O\left( \frac{m^{2}}{M^{2}}\chi
^{4}\right) ,  \label{susy2}
\end{eqnarray}%
where $\phi ^{2}=\phi _{1}^{2}+\phi _{2}^{2},$ $\Delta ^{^{\prime }4}=\Delta
^{4}+16m_{\varphi }^{2}f^{2}+192\lambda _{2} \frac{m^{2}}{M^{2}_M}f^{4}$
and $\alpha =\tan ^{-1}\left( \frac{\phi _{1}}{%
\phi _{2}}\right).$ 
The second term is a $D-$term which only arise when
supersymmetry is broken and hence is proportional to the SUSY breaking
scale, $m^{2}$. It can come from radiative corrections with a messenger
field, $M$, of mass $M_{M}$ in the loop. 

A particularly simple case to analyse and one that allows for the lowest scale of inflation is the case that the hybrid fields have only soft supersymmetry breaking mass, $m_{\chi_i}=O(\Delta^2/M)$. The condition for the end of inflation is 
\begin{equation}
8\lambda _{3}\Delta ^{2}\frac{f^{2}}{M^{2}}\sin \left( \frac{\phi _{e}}{2f}%
\right) \approx m_{\chi i}^{2},  \label{end}
\end{equation}
where $\lambda _{3}$ is a coefficient expected to be of $O(1)$ and $m_{\chi _i}
$ is a soft supersymmetry breaking mass of $O(\frac{\Delta ^{2}}{M})$. Thus in this case, taking $\lambda_3=O(1)$, $\phi_e$ is determined by two of the remaining inflationary parameters 
\begin{equation}
\phi _{e}=O\left( \frac{\Delta ^{2}}{4f}\right) .  \label{phi}
\end{equation}
Since $\phi _{\mathrm{H}}\approx \phi _{e}$ one sees that the
dependence on $\Delta $ cancels in Eq.(\ref{scale}) and the density fluctuations
are determined by $f$. To get the observed density fluctuations
\begin{equation}
4\frac{f}{M}\simeq 10^{-5}.  \label{f}
\end{equation}
Using this Eq.$\left( \ref{fluctuations}\right) $ gives
\begin{equation}
a\simeq \frac{1}{2}\delta _{\mathrm{ns}}\left(\frac{f}{M}\right)^2\simeq 10^{-13} . \label{a}
\end{equation}%
For the supersymmetric model we have from Eq.(\ref{susy2})
\begin{equation}
a=64\lambda _{2}\frac{m^{2}}{M^{2}_M}\frac{f^{4}}{\Delta ^{^{\prime }4}}=64\lambda_2\left(\frac{f}{M}\right)^2 \left(\frac{f}{M_M}\right)^2
,\end{equation}
so Eq.(\ref{a}) can be satisfied by the choice of the messenger mass scale, $f/M_M\approx 10^{-2}/\sqrt{\lambda_2}$.

From Eqs.(\ref{tensor}), (\ref{running}) and Eq.(\ref{f}) we get
\begin{eqnarray}
r & < & 2\,\delta _{\mathrm{ns}}^{2}\,\left( 
\frac{f}{M}\right)^2 \,\simeq  \,6\times 10^{-14}, \label{tensorapp} \\
n_{r} & < & \, \frac{1}{2}\,\delta _{\mathrm{ns}}^{2} \,\simeq\,  1.4\times 10^{-3} .\label{runningapp}
\end{eqnarray}
Since $\Delta^{4}=\frac{1}{8}M^4A_{\mathrm{H}}^{2}r$ we obtain the bound for $\Delta < 1.6\times 10^{13} GeV .$

To summarise, for the case the hybrid fields have only soft supersymmetry breaking masses, fitting the observed values of the spectral index and magnitude of the density perturbations give two constraints on the three hybrid inflation
parameters. The remaining parameter, that can be taken to be the scale of
inflation, $\Delta ,$ is undetermined. Its upper bound is $10^{13}GeV$. 
It has a lower bound because gravitational corrections give an irreducible mass of $O(\bar{\Delta} ^{2}/M)\ $to the $\chi $ field where $\bar{\Delta}$ is the zero temperature supersymmetry breaking scale after inflation has ended. Keeping only this contribution to $m_{\chi i}$ and solving Eq.(\ref{end}) gives 
$
\phi_e=O\left(\bar{\Delta}^4/(4f\Delta^2)\right)
$
and using this in Eq.(\ref{scale}) one finds 
$\Delta^4\approx 10^{-5}\left(\frac{M}{4f}\right)\bar\Delta^4$. To avoid large gravitational corrections we require $f/M<1$ which implies $\Delta>4\times 10^{-2}\bar\Delta$. In
gravity mediated schemes the need to split the superpartners from the
Standard Model states requires $\bar{\Delta }=O(10^{10}GeV)$ but in
gauge mediated schemes $\bar{\Delta}$  can be as low
as $10^{4}GeV$ implying  $\Delta >400GeV$, close to the electroweak breaking scale.

\section{Initial conditions for inflation}
We have argued that hybrid natural inflation with no fine tuning of parameters requires a small value for $\frac{\phi_{\mathrm{H}}}{2f}$. Moreover one may see from Eq.(\ref{phi}) that this ratio becomes very small indeed for the scale of inflation near the lower bound just discussed. This immediately raises the question of initial conditions and how a very small initial value for $\phi$ can be achieved. To answer this question consider the thermal effects present before inflation. In general  $\phi$ will have coupling to other fields with a coupling that is $D_4$ symmetric but not $SU(2)$ symmetric. For example if there is a second doublet field $\psi=\left( 
\begin{array}{c}
\psi _{1} \\ 
\psi _{2}%
\end{array}%
\right) $
there is a $D_4$ invariant superpotential coupling $(\phi_1\psi_1-\phi_2\psi_2)\chi_1^{-+}$ that is not $SU(2)$ invariant. From Eq.(\ref{V}) one may see that this includes a term of the form $\psi_1\chi_1^{-+}\sin(\phi/f)$. Such terms can maintain the fields in thermal equilibrium and generate a term of the form $\sin(\phi/f)^2T^2$ in the effective scalar potential at temperature $T$. This in turn drives the initial value of $\phi$ to be small. 
To quantify this we suppose that at temperature $T_1$ the phase transition associated with the field $\varphi$ occurs. As this occurs the $vev$ of the angular variable $\phi_{T_1}/f$ is undetermined so we expect $\phi_{T_1}/f=O(1)$. The start of inflation occurs at a temperature $T_2$ of $O(\Delta)$ and subsequently the fields fall out of thermal equilibrium. In the intervening period the $vev$ of the field is reduced  $\phi_{T_2}=\phi_{T_1}e^{-m_\phi/H}\approx\phi_{T_1}e^{-M/\Delta}$ where the Hubble parameter $H=T^2/M$ and the final estimate follows using the thermal mass. One sees that such thermal effects readily set the necessary initial conditions for low scale natural hybrid inflation. One may worry that thermal fluctuations $\delta\phi=O(T)$ negates this conclusion, destabilising the hybrid field through the fourth term on the right hand side of Eq.(\ref{susy2}). However this is not the case as it is the mean field values that are relevant to the phase transitions and the thermal fluctuations do not contribute to the mean value of the $\sin(\phi/f)$ term, being odd in $\phi$.

\section{Summary and conclusions}

In summary, we have explored the phenomenological aspects of
\textit{Hybrid Natural Inflation} in which the inflaton is a pseudo Goldstone boson and hence does not suffer from the \textquotedblleft eta\textquotedblright problem. The end of inflation is driven by a hybrid sector and as a result the scale of symmetry breaking associated with the pseudo Goldstone boson inflaton can be smaller than the Planck scale. In contrast with \textit{Natural Inflation} there is no conflict of such a sub-Plankian value with the requirement of generating enough inflation or with bounds imposed by the spectral index. For the supersymmetric model discussed here the inflaton potential relevant at the time density perturbations are produced is governed by three parameters. Two of the parameters are determined by fitting the observed spectral index and the magnitude of density perturbations. The scale of inflation is set by the remaining parameter and can be anywhere in the range $400GeV<\Delta<10^{16}GeV$ and thermal effects can set the initial conditions necessary for low scale inflation provided there is a period of thermal equilibrium after the initial phase transition associated with the Goldstone boson. Thus \textit{Hybrid Natural Inflation} provides another example of inflationary models capable of generating very low scales of inflation. However in contrast with previous examples \cite{German:2001tz} there is no difficulty in fitting the observed spectral index for even the lowest scale of inflation.

\section{Acknowledgements}

The research was partially supported by the EU RTN grant UNILHC 23792. The work by GG is part of the Instituto Avanzado de Cosmolog\'{\i}a (IAC) collaboration.

\end{document}